\documentclass[twocolumn,aps,pra,superscriptaddress,showpacs]{revtex4-1}
\usepackage{epsfig}
\usepackage[colorlinks=true,citecolor=blue,urlcolor=blue,linkcolor=blue]{hyperref}
\usepackage{chemscheme}
\usepackage{graphicx}

\usepackage{multirow}
\usepackage[table,xcdraw]{xcolor}
\usepackage{natbib}
\usepackage{chemfig}
\usepackage[version=4]{mhchem}
\usepackage[normalem]{ulem}
\usepackage{floatrow}
\usepackage{tabulary}
\usepackage{booktabs}
\usepackage[flushleft]{threeparttable}
\usepackage{multirow}
\usepackage{xspace}
\usepackage{xcolor} 
\definecolor{darkgreen}{rgb}{0.0, 0.6, 0.0}
\DeclareMathAlphabet{\mathbsf}{OT1}{cmss}{bx}{n}

\newcommand{\cm}{cm$^{-1}$\xspace}

\usepackage{xspace}

\usepackage{tablefootnote}

\newcommand{\syn}{$syn$}
\newcommand{\anti}{$anti$}
\newcommand{\X}{$\widetilde X$}
\newcommand{\A}{$\widetilde A$\xspace}
\newcommand{\B}{$\widetilde B$}
\newcommand{\AtoX}{ $\widetilde A \rightarrow \widetilde X $}

\newcommand{\BtoX}{ $\widetilde B \rightarrow \widetilde X $}

\usepackage{titlesec}
\titlespacing*{\subsection}
  {0pt}{1\baselineskip}{1\baselineskip}
\usepackage[T1]{fontenc}
\usepackage{threeparttable}

\begin{document}

\title{Witnessing molecular conformation via optical cycling centers}
\affiliation{Department of Physics and Astronomy, University of California, Los Angeles, California 90095, USA}
\affiliation{Department of Chemistry and Biochemistry, University of California, Los Angeles, California 90095, USA}
\affiliation{Center for Quantum Science and Engineering, University of California, Los Angeles, California 90095, USA}
\affiliation{Challenge Institute for Quantum Computation, University of California, Los Angeles, California 90095, USA}

\author{Guo-Zhu Zhu}
\affiliation{Department of Physics and Astronomy, University of California, Los Angeles, California 90095, USA}
\affiliation{Department of Chemistry, Fudan University, Shanghai, 200438, China}

\author{Taras Khvorost}
\affiliation{Department of Chemistry and Biochemistry, University of California, Los Angeles, California 90095, USA}

\author{Haowen Zhou}
\affiliation{Department of Physics and Astronomy, University of California, Los Angeles, California 90095, USA}

\author{Guanming Lao}
\affiliation{Department of Physics and Astronomy, University of California, Los Angeles, California 90095, USA}

\author{Claire E. Dickerson}
\affiliation{Department of Chemistry and Biochemistry, University of California, Los Angeles, California 90095, USA}

\author{Cecilia Chang}
\affiliation{Department of Chemistry and Biochemistry, University of California, Los Angeles, California 90095, USA}

\author{Ryan Choi}
\affiliation{Department of Chemistry and Biochemistry, University of California, Los Angeles, California 90095, USA}

\author{Wesley C. Campbell}
\affiliation{Department of Physics and Astronomy, University of California, Los Angeles, California 90095, USA}
\affiliation{Center for Quantum Science and Engineering, University of California, Los Angeles, California 90095, USA}
\affiliation{Challenge Institute for Quantum Computation, University of California, Los Angeles, California 90095, USA}

\author{Anastassia N. Alexandrova}
\affiliation{Department of Chemistry and Biochemistry, University of California, Los Angeles, California 90095, USA}
\affiliation{Center for Quantum Science and Engineering, University of California, Los Angeles, California 90095, USA}

\author{Eric R. Hudson}
\affiliation{Department of Physics and Astronomy, University of California, Los Angeles, California 90095, USA}
\affiliation{Center for Quantum Science and Engineering, University of California, Los Angeles, California 90095, USA}
\affiliation{Challenge Institute for Quantum Computation, University of California, Los Angeles, California 90095, USA}

\date{\today}

\begin{abstract}
Molecular conformation plays a central role in determining molecular properties and dynamics. 
Here, we report the production and conformer-resolved spectroscopy of cold calcium 3-hydroxyphenoxide radical (CaOC$_6$H$_4$-3-OH) in a cryogenic buffer-gas cell. Internal rotation of the hydroxyl group gives rise to distinct \anti~and \syn~conformers. The CaO moiety acts as an optical cycling center (OCC), enabling the $\widetilde A - \widetilde X$ and $\widetilde B - \widetilde X$ transitions of both conformers to be resolved by laser-induced fluorescence spectroscopy. Dispersed fluorescence measurements reveal diagonal vibrational branching fractions of 0.91-0.93, indicating favorable optical cycling characteristics. Following  excitation of off-diagonal vibronic transitions near the calculated torsional barrier in the \A~state, the emission remains predominantly conformer specific, indicating that conformational identity is largely preserved over the excited-state fluorescence lifetime. These results demonstrate that a metal-centered OCC can serve as an intrinsic, site-specific optical reporter of molecular conformation, providing a foundation for repetitive optical readout and studies of conformational dynamics at the single-molecule level.
\end{abstract}

\maketitle

\section*{INTRODUCTION}
Molecular conformers are distinct spatial arrangements of a molecule with identical atomic connectivity. Their structural differences influence chemical reactivity and molecular recognition, leading to distinct spectroscopic signatures.\cite{canfield2018new,kuhlman2019advances} Conformers have therefore been widely investigated in spectroscopy,\cite{de2007spectroscopy,rizzo2009spectroscopic,park2021development} reaction dynamics \cite{chang2013specific,champenois2021conformer,park2002photodissociation,kim2007conformationally}, and intramolecular tunneling.\cite{mardyukov2017conformer,fausto2022ir,pettersson2003conformational} Nevertheless, the selective identification, preparation, and control of individual conformers remain challenging because small torsional barriers can allow rapid interconversion at ambient temperature. Low-temperature techniques, such as cryogenic matrix isolation \cite{mardyukov2017conformer,fausto2022ir,pettersson2003conformational}, supersonic beam expansion\cite{park2021development,dian2004direct}, buffer-gas cooling \cite{rizzo2009spectroscopic,satterthwaite2023low,borengasser2023conformer}, and sympathetic cooling of ions \cite{chang2013specific,xu2024trapping}, can suppress thermal interconversion, enabling conformer-resolved spectroscopic characterization and even selective preparation. Despite these advances, many existing methods rely on destructive detection schemes, such as photoionization \cite{park2021development} or photodissociation \cite{park2002photodissociation,kim2007conformationally}. While infrared \cite{fausto2022ir,mardyukov2017conformer} and microwave spectroscopy \cite{dian2008measuring,hernandez2017conformational} can probe conformers non-destructively, their relatively small absorption cross-sections limit the sensitive, repetitive readout required for quantum state manipulation at the single-molecule level. A bright, intrinsic optical readout has yet to be realized for conformational isomers. 

Optical cycling centers (OCCs) provide a promising route toward such an optical
interface. 
An OCC can often be realized from an alkaline-earth metal atom, such as Ca or Sr, attached to the molecular framework either directly or through an electronegative atom, as in CaO$-$ and SrO$-$.\cite{Isaev2016Polyatomic,kozyryev2016proposal,zhou2025vibronic} When incorporated into molecules, the unpaired radical electron remains strongly localized on the metal center, giving rise to atomic-like electronic structures. The resulting minimal change in molecular geometry upon excitation leads to highly diagonal Franck-Condon factors (FCFs) and nearly
closed optical cycling transitions. These properties enable repeated photon scattering and have advanced laser cooling and trapping of diatomic molecules, such as SrF~\cite{shuman2010laser}, CaF~\cite{anderegg2017radio}, BaF~\cite{zeng2024BaF}, AlF~\cite{padilla2025AlF}, and small polyatomic molecules including SrOH~\cite{lasner2025sroh}, CaOH~\cite{vilas2022CAOH}, and CaOCH$_3$~\cite{mitra2020direct}. The functionalization of OCC moiety on large polyatomic molecules has been proposed theoretically \cite{Isaev2016Polyatomic,kozyryev2016proposal,augenbraun2020molecular,ivanov2020two,ivanov2020toward,klos2020prospects,Dickerson2021FranckCondon,dickerson2021optical,dickerson2022fully,yu2023multivalent,khvorost2024dual,Wjcik2024}
and demonstrated spectroscopically for a series of alkoxides,\cite{paul2021electronic,telfah2022combined} aromatic
compounds~\cite{mitra2022pathway,augenbraun2022high,zhu2022functionalizing,lao2022sroph,zhu2024extending}, and adamantane-based derivatives,\cite{Lao2025} thereby extending the potential of optical cycling to organic systems and
opening new frontiers in precision measurement and quantum information science.\cite{augenbraun2023direct,cornish2024quantum,demille2024quantum}

The optical signature of an OCC serves not only as a handle for quantum control but also as a sensitive probe of its local electronic environment. In aromatic systems of MOPh-X (M = Ca or Sr, Ph = phenyl, X = substituent), ring substituents modulate the metal-centered transition frequencies and FCFs through inductive and resonance effects, allowing, in principle, an OCC to report detailed chemical information about its ligands. \cite{Dickerson2021FranckCondon,dickerson2021optical,zhu2022functionalizing,lao2022sroph,zhu2024extending,Lao2025} However, a fundamentally different and as-yet-unexplored question remains: can an OCC distinguish between two conformations of the same molecule, where the perturbation to the metal center's electronic environment arises solely from a change in the spatial orientation of a functional group, rather than from a change in chemical identity? 

\begin{figure}
    \centering
    \includegraphics[scale=0.5]{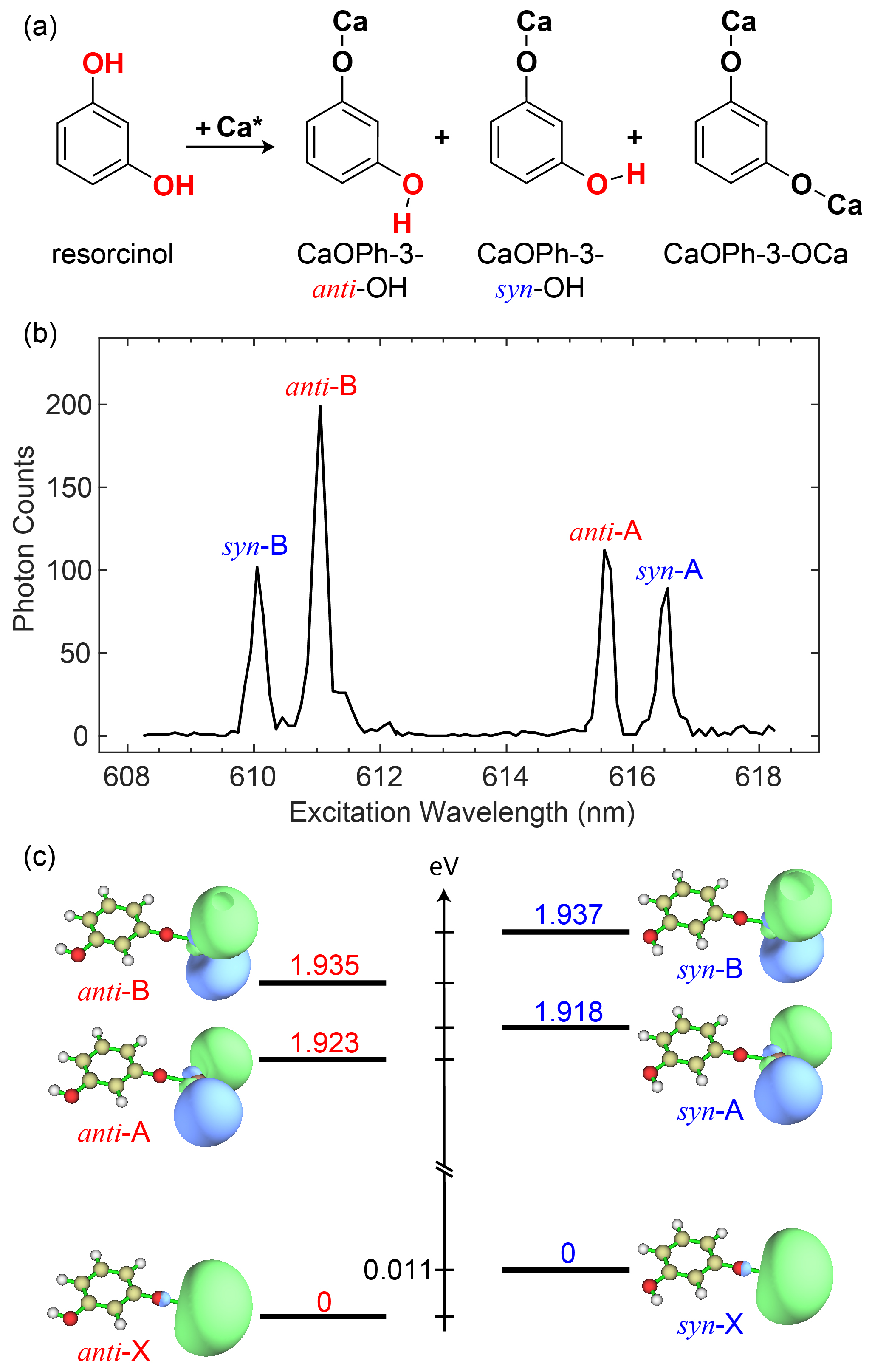}
    \caption{(a) The reaction of resorcinol with metastable excited Ca atoms can produce $single-$OCC molecules with two conformations of CaOPh-3-\syn-OH and CaOPh-3-\anti-OH, and $bi-$OCC molecule CaOPh-3-OCa. (b) Excitation scan of the reaction products shows four peaks that are respectively assigned to \syn-B (i.e., $\widetilde B - \widetilde X$ of CaOPh-3-\syn-OH), \anti-B, \anti-A and \syn-A. (c) Calculated energy levels of \X, \A and \B~levels of both conformers and the corresponding natural transition orbitals of CaOPh-3-OH. \anti-X has a
    lower energy of 11 meV than \syn-X.  Calculated transition energies are given.
    } 
    \label{fig:structure}
\end{figure}

Here, we address this question using calcium 3-hydroxyphenoxide (CaOPh-3-OH) as a model system. As shown in Fig. \ref{fig:structure}a, the precursor resorcinol (HOPh-3-OH) reacts with Ca atoms to form CaOPh-3-OH (See SI for experimental details). The hydroxyl substituent at the \textit{meta} position gives rise to \anti~and \syn~conformers, defined by its orientation relative to the Ca-O axis and separated by a calculated torsional barrier of 0.15 eV.  
Using laser-induced fluorescence (LIF) excitation spectroscopy, we demonstrate that the OCC
unambiguously distinguishes the two conformers through their distinct $\widetilde A - \widetilde X$ and $\widetilde B - \widetilde X$ transitions, while dispersed LIF (DLIF) spectroscopy further reveals favorable vibrational branching fractions (VBFs) for both conformers. The resolved conformer-specific transitions and favorable VBFs support the development of a bright, intrinsic, and potentially repetitive optical readout of conformational states, an important step toward quantum state manipulation of individual molecules. We also use off-diagonal excitation to access vibrational levels with energies approaching the torsional barrier in the \A state and use conformer-specific fluorescence to search for vibrationally induced interconversion. These measurements assess the capability of a metal-centered OCC to distinguish conformational states while retaining favorable optical cycling properties and to serve as an optical probe of remote structural dynamics.

\begin{figure*}
    \centering
    \includegraphics[scale=0.39]{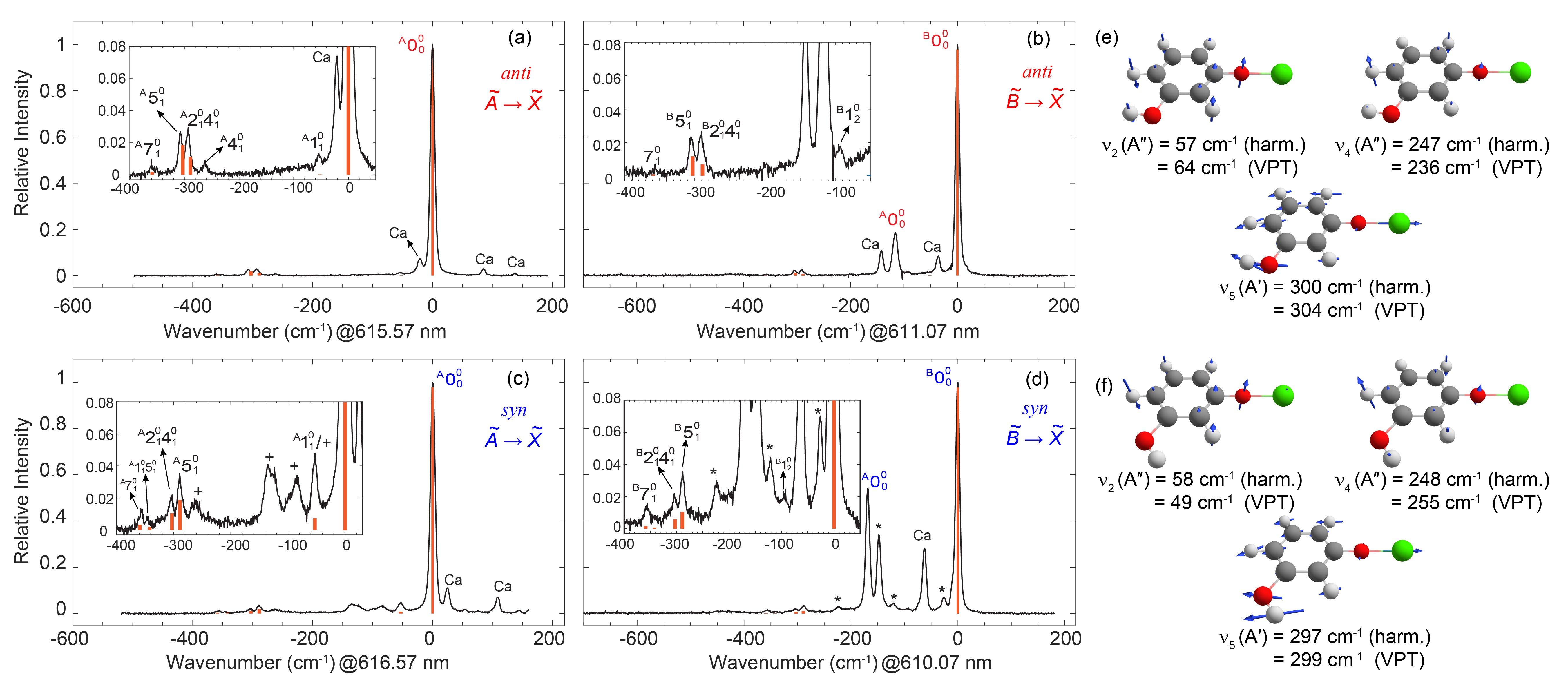}
    \caption{DLIF spectra of CaOPh-3-OH species. (a) \anti~\AtoX, (b) \anti~\BtoX, (c) \syn~\AtoX, and (d) \syn~\BtoX. The vertical orange lines indicate the theoretical frequencies and FCFs in VPT anharmonic calculations. The identified vibrational decay channels are labeled as $^{A/B}\nu^0_i$, where $A$ or $B$ denotes the upper electronic state, $\nu_i$ denotes the $i$th vibrational mode populated in the \X~state, and the superscript $0$ indicates that the decay originates from the vibrational ground level of the upper state. The peaks labeled with Ca represent the Ca atomic lines at 610 nm ($4s5s~^3S_1 \rightarrow
    4s4p~^3P_0$), 612 nm ($4s5s~^3S_1 \rightarrow 4s4p~^3P_1$) and 616 nm
    ($4s5s~^3S_1 \rightarrow 4s4p~^3P_2$), respectively. They are produced by laser ablation. The peaks labeled with + indicate the contamination from CaOH, while those with * are unknown contamination. (e) and (f) show calculated normal mode displacement vectors involved in the observed $\nu_5 \leftrightarrow \nu_2 + \nu_4$ Fermi resonance coupling for the \anti~and \syn~conformers, respectively. The corresponding harmonic and VPT anharmonic frequencies are given.
    } 
    \label{fig:ca-dlif}
\end{figure*}

\section*{RESULTS AND DISCUSSION} 

\noindent{\textbf{Conformer-resolved excitation spectroscopy and OCC-based conformational sensing.}}
The reaction of metastable excited Ca atoms  with resorcinol in a cryogenic buffer gas cell can produce three species: the \anti~and \syn~conformers of the $single-$OCC molecule (CaOPh-3-OH), and a $bi-$OCC product (CaOPh-3-OCa), as illustrated in Fig. \ref{fig:structure}a. The LIF excitation spectrum of the reaction products in the cryogenic cell shows four well-resolved peaks in the wavelength range of 608-618 nm (Fig. \ref{fig:structure}b).~

To assign these features, we performed electronic structure calculations for both conformers of the $single-$OCC species, considering the two lowest electronically excited states, \A and \B~ (See SI for computational details). The results, summarized schematically in Figs. \ref{fig:structure}c, provide two complementary pieces of evidence for the assignment. First, the \anti~conformer is calculated to lie 11 meV below the \syn~conformer in ground-state energy. Its stronger LIF signals are qualitatively consistent with a larger population. However, thermal equilibrium at 20 K would strongly suppress the \syn~population. The clear observation of the \syn~conformer therefore indicates a kinetically trapped, nonequilibrium distribution during molecular formation and buffer-gas cooling. Second, the calculated excitation energies increase in the order of \syn-A (1.918 eV) $<$ \anti-A (1.923 eV) $<$ \anti-B (1.935 eV) $<$ \syn-B (1.937 eV), placing the two \anti~conformer transitions between those of the \syn~conformer. Taken together, these results readily assign the four peaks in Fig. \ref{fig:structure}b, in order of increasing excitation energy, to \syn-A, \anti-A, \anti-B and \syn-B, respectively. The measured \A--\B~splittings are $\approx$ 15 meV and 21 meV for the \anti~and \syn~conformers, respectively, in good agreement with the corresponding calculated values of 12 meV and 19 meV. For both conformers, the natural transition orbitals (NTOs) involved in these transitions are predominantly localized on the Ca atom, with $s-$type character in the ground state and $p-$type character in the two excited states, consistent with previous studies. \cite{Dickerson2021FranckCondon,dickerson2021optical,dickerson2022fully,mitra2022pathway,augenbraun2022high,zhu2022functionalizing,lao2022sroph,zhu2024extending,lao2022sroph}
No transitions attributable to the $bi-$OCC species are observed in the excitation spectrum. 
This absence is likely due to a low formation yield under our experimental conditions, as formation requires either secondary reactions, involving metastable metal atoms, or three-body collisions. \cite{khvorost2024dual,brudner2025insights}

These results establish the OCC as a local optical reporter capable of distinguishing conformers separated by a small interconversion barrier based on their resolved electronic transitions. Additional calculations on naphthalene- and pyrene-based analogues indicate that the calculated separation between the \anti-A and \syn-A transition energies decreases as the rotating OH group is positioned farther from the OCC (Fig. \ref{fig:TDDFT-fig}), suggesting that the conformational sensitivity of the OCC is predominantly local. 

\begin{figure*}
    \centering
    \includegraphics[scale=0.6]{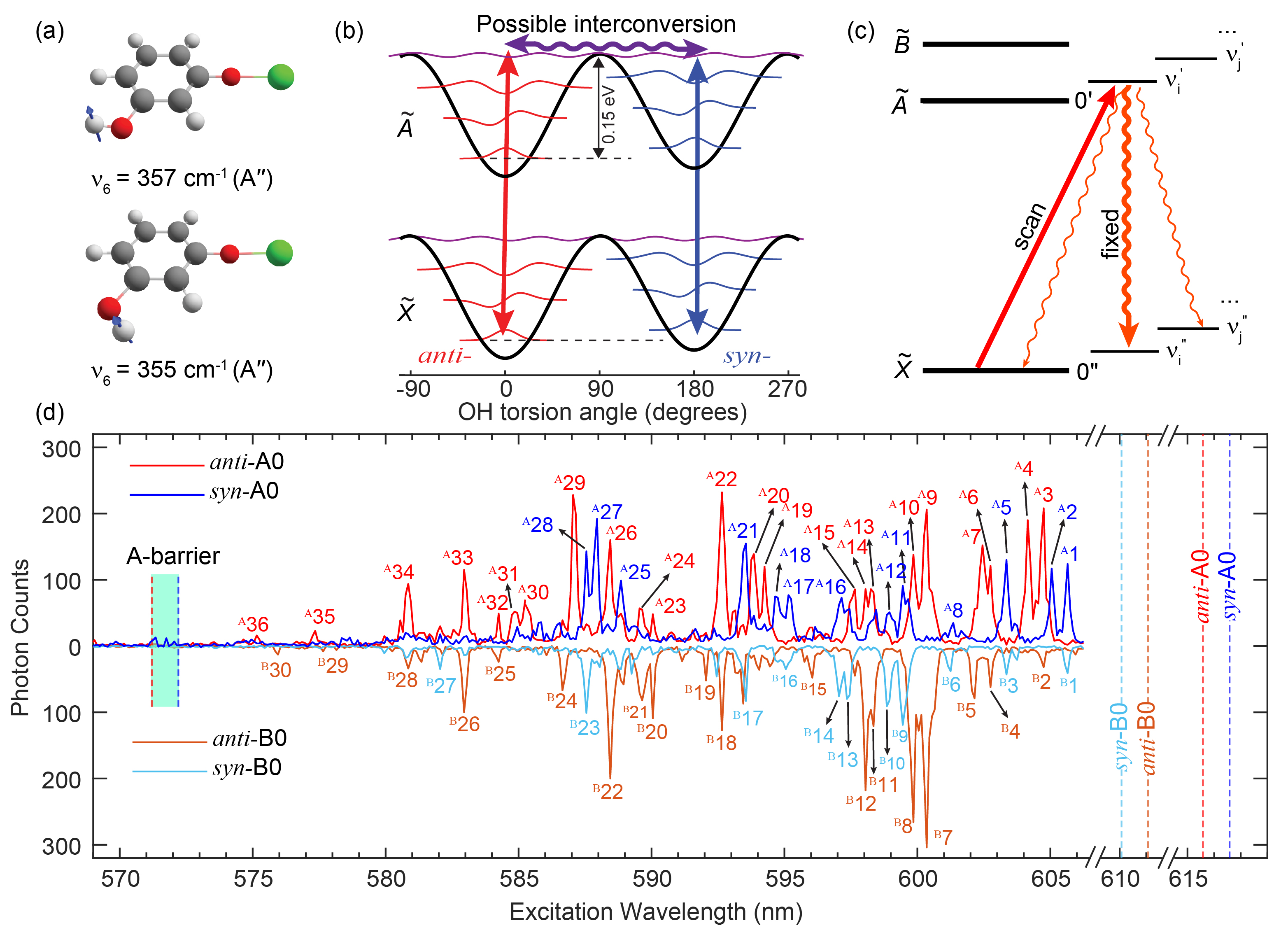}
    \caption{(a) Vibrational displacements of $\nu_6$ torsional mode of both conformers. Both modes involve the OH flipping responsible for the conformational switching. The calculated harmonic frequencies are given. (b) Schematic periodic potentials along the OH torsional coordinate in the \X~and \A~states. The endpoints at $-90^\circ$ and $270^\circ$ represent the same geometry, giving rotational paths between \anti~and \syn. The calculated rotation barrier of 0.15 eV is given for the \A~state of the \anti~conformer. The colored curves illustrate vibrational states schematically and are not calculated torsional wavefunctions. The vertical double-headed arrows indicate conformer-specific excitation and fluorescence; the horizontal arrow indicates possible excited-state interconversion. (c) Excitation and detection scheme for the off-diagonal excitation scans. The excitation wavelength is scanned across vibrational excited levels, while fluorescence is monitored at a fixed wavelength near the corresponding vibrationally diagonal emission. (d) Full excitation spectra of the $\widetilde A (\nu_n')\leftarrow \widetilde X (\nu''=0)$ (upward) and $\widetilde B (\nu_n')\leftarrow \widetilde X (\nu''=0)$ (downward) of both conformers. The emission fluorescence is monitored on respective diagonal decays at 615.2 nm (\anti~\A), 616.1 nm (\syn~\A), 610.7 nm (\anti~\B) and 609.6 nm (\syn~\B). The dash lines indicate the excitation wavelengths of the diagonal 0-0 transitions. The shaded region at 572 nm represents the calculated torsional barrier in the \A~state. 
    } 
    \label{fig:ca-pes-ex}
\end{figure*}

\noindent{\textbf{Conformer-resolved vibrational branching and optical-cycling prospects.}}
Having established conformer-resolved optical signatures, we next characterized the vibrational branching of each conformer to assess their optical cycling properties. The excitation laser wavelengths were tuned to each of the four peaks in Fig. \ref{fig:structure}b, and DLIF spectra were recorded for radiative decay from the $\widetilde A$ and $\widetilde B $ states of both conformers (Fig. \ref{fig:ca-dlif}). For display, the spectra are plotted as wavenumber shifts relative to the excitation frequency and normalized to their respective 0-0 peaks. All four spectra are dominated by sharp diagonal 0-0 peaks, with only weak off-diagonal vibrational features. The vibrational bands were assigned by comparing the observed shifts with calculated harmonic and vibrational-perturbation-theory (VPT) anharmonic frequencies \cite{Boyer_2022,zhu2024extending} and FCFs (Tables \ref{tab:theo-freqs}--\ref{tab:theo-fcf}) at the PBE0/def2-TZVPPD level of theory. Vibrational branching fractions were determined from the integrated areas obtained by fitting the bands with Voigt profiles.\cite{zhu2022functionalizing,lao2022sroph,zhu2024extending,Lao2025} 

Figure \ref{fig:ca-dlif}a presents the dispersed spectrum following excitation of the $\widetilde A - \widetilde X$ origin band of the \anti~conformer at 615.57 nm. The most prominent off-diagonal features appear near $-$300 \cm, and a magnified view shows two resolved peaks. Based on the  calculated vibrational frequencies and FCFs (Tables \ref{tab:theo-freqs}--\ref{tab:theo-fcf}), the peak at $-$304 \cm is assigned to the $\widetilde{A}(v'=0)\rightarrow\widetilde{X}(\nu_5=1)$ transition, where $\nu_5$ is an in-plane Ca-O/ring stretching mode (Fig.\ref{fig:ca-dlif}e). The neighboring peak at $-$290 \cm is assigned to the combination band of two out-of-plane bending modes $\nu_2 + \nu_4$ (Fig.\ref{fig:ca-dlif}e). It gains intensity through Fermi resonance mixing with $\nu_5$ \cite{zhu2024extending}, as predicted by VPT calculations, although the calculation gives the opposite relative intensity ordering. A weaker band at $-$357 \cm is assigned to another in-plane Ca-O/O-H scissoring mode $\nu_7$ (Fig. \ref{fig:vib-modes}), as predicted by the calculations. Two additional weak bands at $-$55 \cm and $-$260 \cm, which are not predicted to carry appreciable intensity by either harmonic or VPT calculations, are tentatively assigned to the fundamental modes $\nu_1$ and $\nu_4$ (Fig. \ref{fig:vib-modes}), respectively, based on their calculated frequencies. Features labeled "Ca" originate from atomic Ca emissions. 

\begin{figure*}
    \centering
    \includegraphics[scale=0.55]{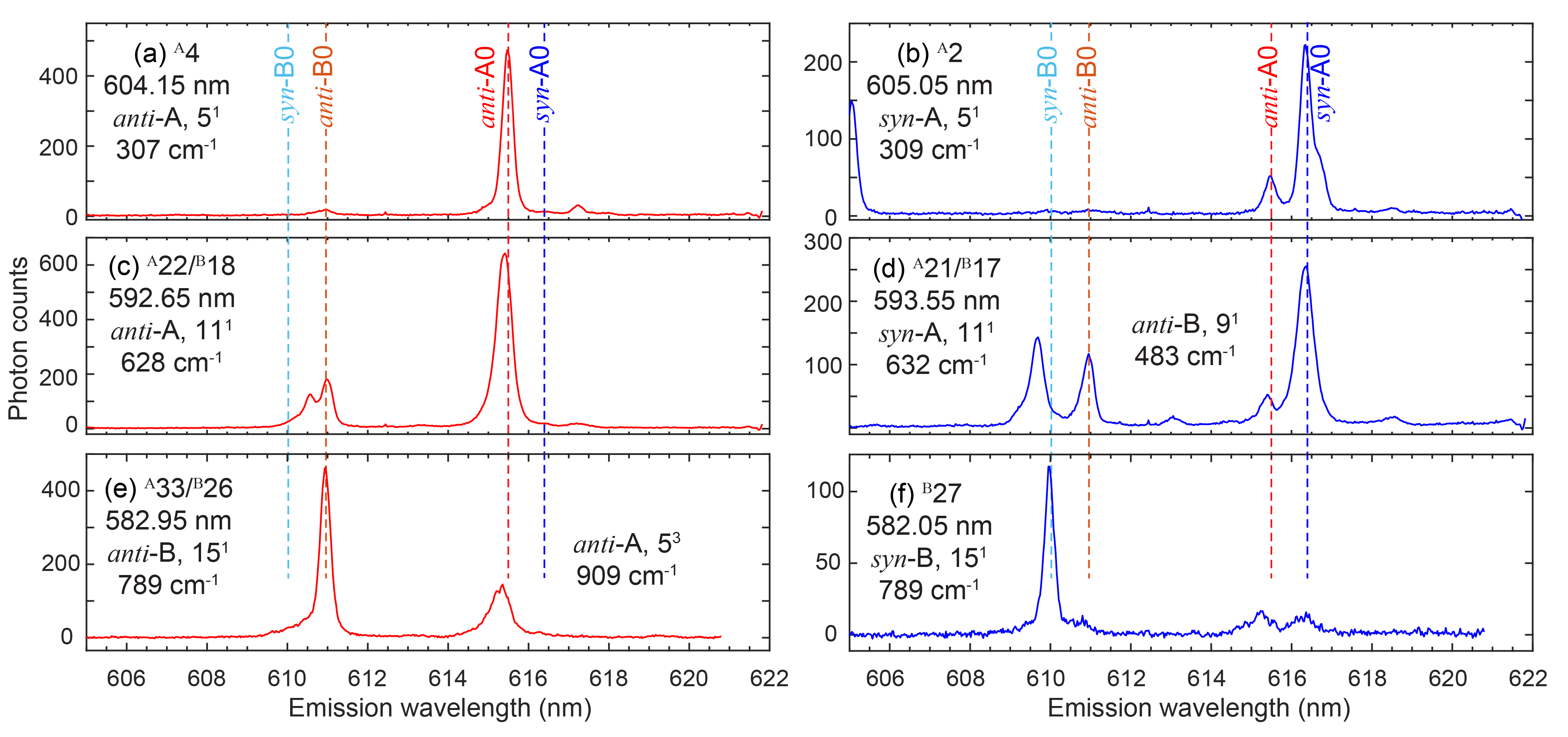}  
    \caption{Six representative DLIF spectra recorded following excitations of peaks (a) $^A4$, (b) $^A2$, (c) $^A22/^B18$, (d) $^A21/^B17$, (e) $^A33/^B26$, and (f) $^B27$ in Fig. \ref{fig:ca-pes-ex}d. Each spectrum is dominated by vibrationally diagonal emission associated with one of the four conformer-specific electronic origins: \anti-A0, \syn-A0, \anti-B0, or \syn-B0. The excitation energy offsets relative to the corresponding electronic origins and the corresponding assignments are indicated in each figure. Assignments of the excitation resonances are obtained by comparing those offsets with the calculated vibrational frequencies (Table. \ref{tab:AB_freq}). 
    } 
    \label{fig:ca-dlif-AX}
\end{figure*} 

The DLIF spectrum of the \anti \BtoX~transition is shown in Fig. \ref{fig:ca-dlif}b. Three prominent peaks within $-$200 \cm of the origin are attributed to either atomic Ca emissions or the \A origin band ($^A0^0_0$) \cite{zhu2022functionalizing,zhou2025PRR}. The inset shows two weak peaks near $-300$ \cm, which are assigned
to $\nu_5$ (Fig.\ref{fig:ca-dlif}c) at $-$302 \cm and $\nu_2 + \nu_4$ combination band at $-288$ \cm, respectively. As in the \AtoX~spectrum in Fig.\ref{fig:ca-dlif}a, the combination band gains intensity due to Fermi resonance
coupling with $\nu_5$. Two additional weak features at $-$92 \cm and $-$354 \cm are assigned to the overtone of in-plane rocking mode 2$\nu_1$ and the fundamental in-plane scissoring mode $\nu_7$ (Fig.\ref{fig:vib-modes}), respectively.

The spectra of the \syn~conformer (Figs.\ref{fig:ca-dlif}c-d) are slightly more congested because of overlapping emissions from atomic Ca and CaOH ("+"). After excluding these contaminant features, the remaining vibrational features closely resemble those of the \anti~conformer. In the \AtoX~spectrum, the two bands near $-300$ \cm are assigned to the same $\nu_5$/($\nu_2 + \nu_4$) Fermi dyad (Fig.\ref{fig:ca-dlif}f). Unlike in the \anti~conformer, the dominant $\nu_5$ lies closer to the origin and is more intense, consistent with VPT calculations. The calculated Hamiltonians and eigenstate compositions (Tables \ref{tab:fermi_hamiltonians} and \ref{tab:fermi_compositions}) show that the exchange of dominant characters results from the reversed deperturbed energy ordering, while the similar coupling strengths and mixing coefficients indicate essentially equivalent Fermi resonance mixing in both conformers. Two weak features near $-354$ \cm corresponds to $\nu_7$ and the $\nu_1 + \nu_5$ combination band, as predicted by the VPT calculations (Table \ref{tab:theo-fcf}). The feature at $-51$ \cm, assigned to the fundamental mode $\nu_1$ (Fig. \ref{fig:vib-modes}), overlaps substantially with the CaOH contamination. The \BtoX~spectrum (Fig.\ref{fig:ca-dlif}d) similarly exhibits the characteristic bands associated with $\nu_5$, $\nu_2 + \nu_4$ and $\nu_7$. An additional feature at $-95$ is tentatively assigned to the overtone 2$\nu_1$. Features marked with asterisk do not match any of the calculated frequencies and likely come from unidentified contaminants. 

Overall, both conformers exhibit vibrational branching dominated by the diagonal 0-0 transition. After excluding the identified contaminant contributions, the fitted areas of the observed peaks yield 0-0 branching fractions of 0.91-0.93, with the remaining intensity distributed among a small number of vibrational channels (Table \ref{tab:exp_vbr}). These approximate values do not include systematic uncertainties including wavelength-dependent instrument response, or unresolved and unobserved decay channels.\cite{zhu2022functionalizing,lao2022sroph,zhu2024extending,Lao2025} Nevertheless, they are comparable to those reported for other calcium phenoxides derivatives \cite{zhu2022functionalizing,lao2022sroph,zhu2024extending,Lao2025} and support conformer-selective fluorescence detection and promising optical cycling. 

\noindent{\textbf{Off-diagonal excitation scan and search for conformational switching.} 
The favorable diagonal FCFs of the two conformers suggest efficient optical cycling, but also raise the question of whether vibrational excitation in the upper electronic state can induce conformational interconversion. \cite{dian2004direct,dian2008measuring,mardyukov2017conformer,pettersson2003conformational,fausto2022ir} Both conformers possess a low-frequency out-of-plane OH rocking mode $\nu_6$ associated with the torsional coordinate (Fig. \ref{fig:ca-pes-ex}a). The calculated torsional barrier in the \A~state is 0.15 eV relative to the \anti~conformer (Fig. \ref{fig:ca-pes-ex}b). Interconversion may occur through tunneling below the barrier or through over-barrier motion when sufficient energy is deposited into, or efficiently coupled to, the torsional coordinate. \cite{fausto2022ir,pettersson2003conformational,dian2008measuring} 
The OH group can rotate from \anti~to \syn~ by passing on either side of the molecular plane. Interference between these two paths may affect whether torsional states near or above the barrier are localized around one conformer or spread across both. To probe the possible interconversion, we performed off-diagonal excitation scans (Fig. \ref{fig:ca-pes-ex}c) by tuning the laser wavelength to vibrationally excited levels of the \A~and \B~states ($\widetilde A/\widetilde B, \nu_n' \leftarrow \widetilde X, \nu''=0$) while monitoring fluorescence within a spectral window around the corresponding 0-0 transition. Because the frequencies of corresponding vibrational modes are similar in the ground (Table \ref{tab:theo-freqs}) and excited electronic states (Table \ref{tab:AB_freq}), the vibrationally diagonal emission ($\widetilde A/\widetilde B, \nu_n' \rightarrow \widetilde X, \nu''_n$) occurs close to the 0-0 transition and falls within this detection window. In this scheme, if conformational switching occurs, excitation of one conformer would produce emission signature of the other conformer. Figure \ref{fig:ca-pes-ex}d shows excitation spectra from 606 to 570 nm, extending to $\approx$ 1300 \cm~above the \syn-A0 and \anti-A0 origins and beyond the calculated torsional barrier in the \A~state near 572 nm.
More than thirty vibrational resonances are observed across the \A~(upward) and \B~(downward) spectra of both conformers. These resonances can be broadly categorized into three types: conformer-specific features, like $^{A}1/^{B}1$ and $^{A}9/^{B}7$; partially overlapped features with contributions to both conformers, such as $^{A}2/^{A}3$ and $^{A}10/^{A}11$; and fully overlapped peaks with distinct intensity ratios, exemplified by $^{A}33/^{B}26$. At higher excitation energies, a larger fraction of features contain contributions from both conformers, reflecting the greater vibrational density of states and resulting spectral congestion.\cite{zhou2025PRR,augenbraun2023direct} In addition, the overall fluorescence signal decreases rapidly at excitation wavelengths below 580 nm, consistent with the calculated very small FCFs for transitions to higher-lying vibrational levels (Table \ref{tab:theo-fcf}). 

The vibrational resonances were tentatively assigned by comparing their excitation-energy offsets with calculated harmonic frequencies and combination energies (Table \ref{tab:AB_freq}). The excitation scans contain several candidate Fermi dyads. 
For example, peaks $^A1/^A2$, located 292/309 \cm above \syn-A0, and peaks $^A3/^A4$, located 291/307 \cm above \anti-A0, are tentatively assigned to the lower and upper members of the ($\nu_2 + \nu_4$)/$\nu_5$ Fermi dyad in the \A~state. The individual vibrational labels remain tentative in the absence of excited-state VPT calculations. All assignments are summarized in Table \ref{tab:ex_peaks_assign}.

To identify the contributing conformers through their characteristic vibrationally diagonal emission, DLIF spectra were recorded for twenty-one  resonances selected to cover both electronic states and conformers, a broad range of excitation energies, and both isolated and spectrally overlapping features. Six representative spectra are shown in Figs. \ref{fig:ca-dlif-AX}, while the remaining spectra are provided in Fig. \ref{fig:SI-Ca-emission}. Figs. \ref{fig:ca-dlif-AX}a and \ref{fig:ca-dlif-AX}b show DLIF spectra following excitations of the upper members $^A4$ (604.15 nm) and $^A2$ (605.05 nm), respectively. Both spectra exhibit dominant emissions near the respective \anti-A0 and \syn-A0, confirming their associations with the corresponding conformers. A weaker \anti-A0 feature in Fig. \ref{fig:ca-dlif-AX}b arises from coincident excitation of the overlapping resonance $^A3$ (Fig. \ref{fig:ca-pes-ex}d), which is assigned to the \anti-A, $2^14^1$ based on its frequency matching and DLIF spectrum in Fig. \ref{fig:SI-Ca-emission}. In Fig. \ref{fig:ca-dlif-AX}c, the excitation-energy offset of 628 \cm and the dominant \anti-A0 emission support the assignment to \anti-A, $11^1$. Fig. \ref{fig:ca-dlif-AX}d is more complex because the excitation wavelength overlaps resonances $^A21/^B17$ and $^A20$ in Fig. \ref{fig:ca-pes-ex}d. Its dominant \syn-A0 emission is assigned to $\nu_{11}$ of the \syn-A state, while the weaker feature near the \anti-A0/B0 is attributed to excitation of the \anti-A, $5^2$ level. The dominant \anti-B0 and \syn-B0 emissions in Figs. \ref{fig:ca-dlif-AX}e and \ref{fig:ca-dlif-AX}f, respectively, are attributed to decays from vibrational level $15^1$ of the corresponding \B~states. 

For all resonances examined by DLIF, the emission remains dominated by the conformer associated with the assigned excitation transition, and no unambiguous evidence for conformational interconversion is observed within the excited-state fluorescence lifetime. This result is consistent with the metal-localized character of the OCC transition (Fig. \ref{fig:structure}c), resulting in weak coupling to the remote OH torsional coordinate and negligible calculated FCF for the torsional mode (Table \ref{tab:theo-fcf}). Consequently, the vibrational levels accessed in the upper electronic states may contain insufficient torsional character to promote measurable switching, even when their total energies approach the calculated barrier. Nevertheless, weak interconversion through sub-barrier tunneling or vibrational coupling to the torsional coordinate cannot be excluded. 

\section*{CONCLUSION AND OUTLOOK} 
In summary, we report the conformer-resolved optical spectroscopy of cold CaOPh-3-OH radicals. LIF excitation spectroscopy resolves four conformer-specific electronic origins, assigned to the $\widetilde A - \widetilde X$ and $\widetilde B - \widetilde X$ transitions of the \anti~and \syn~conformers. These distinct transitions enable conformer-selective excitation and spectroscopic identification of each conformer. This finding establishes that an OCC can distinguish conformers that differ solely by the spatial orientation of a remote hydroxyl group. DLIF measurements further yield highly diagonal vibrational branching fractions of 0.91-0.93 for all four transitions, indicating that favorable optical cycling properties are retained in both conformers. Off-diagonal excitation scans reveal an extensive manifold of excited vibrational levels in the \A~and \B~states, extending to energies near the calculated torsional barrier in the \A state. For all levels examined by DLIF, however, the emission remains dominated by the conformer associated with the initially excited transition, and no conclusive evidence for conformational interconversion within the excited-state lifetime is observed.

Together with the recent proposal to use an OCC to probe photoisomerization in azobenzene derivatives,\cite{Wjcik2024} our findings suggest a broader role for OCCs as local optical reporters of molecular structure and dynamics. The sensitivity of the OCC transition to a remote structural change, combined with its favorable vibrational branching, could enable repeated optical readout of molecular conformation. This capability could facilitate studies of photochemical transformations,\cite{Wjcik2024} non-adiabatic couplings \cite{zhou2025PRR,wojcik2026unraveling} and anharmonic vibrational coupling and energy redistribution.\cite{zhu2024extending,augenbraun2023direct} Ultimately, molecules incorporating OCCs could provide a platform for the independent control and optical readout of remote structural dynamics in individual molecules. 

\emph{Acknowledgements --} This work was supported by the AFOSR (grant no. FA9550-20-1-0323), the NSF (grant no. OMA-2016245, PHY-2207985, PHY-2110421 and DGE-2034835), NSF Center for Chemical Innovation Phase I (grant no. CHE-2221453). This research is funded in part by the Gordon and Betty Moore Foundation. Computational resources were provided by XSEDE and UCLA IDRE shared cluster hoffman2. This work also used Bridges-2 at the Pittsburgh Supercomputing Center through allocation CHE230091 from the Advanced Cyberinfrastructure Coordination Ecosystem: Services \& Support (ACCESS) program, which is supported by U.S. National Science Foundation grants 2138259, 2138286, 2138307, 2137603, and 2138296.

\bibliographystyle{apsrev4-1_no_Arxiv}
\bibliography{ref}

\onecolumngrid

\section{Supplementary Information}

\subsection{Experimental methods}
The experiments were conducted within a cryogenic buffer-gas cell operated maintained at 20-25 K.\cite{lao2022sroph,zhu2024extending,Lao2025} CaOPh-3-OH radicals were produced by reacting Ca atoms with resorcinol (Sigma Aldrich). An Nd:YAG laser (Minilite) operating at 1064 nm with a pulse energy of approximately 6 mJ and a repetition rate of 10 Hz, was used to ablate calcium metal pellet. To minimize variations in molecular production, the focused ablation spot was continuously swept across the target using a movable mirror. The ablated Ca atoms reacted with resorcinol vapor delivered from a heated reservoir through a heated gas line. The resulting molecules were collisionally cooled by neon buffer gas with a number density of approximately 10$^{15-16} cm^{-3}$. In the excitation region, a tunable pulsed dye laser (PDL, LiopStar-E), operating at 10 Hz with a linewidth of 0.04 \cm~at 620 nm, was utilized to excite the electronic transitions of CaOPh-3-OH. The resulting fluorescence, with a lifetime of approximately 20-30 ns,\cite{zhu2022functionalizing,mitra2022pathway} was collected using a lens system and directed into a monochromator (McPherson Model 2035) equipped with a 1200 lines/mm grating. All excitation scans were recorded using a photomultiplier tube (PMT), with photon counts accumulated by an SRS SR430 multichannel scaler, whereas the DLIF spectra were recorded using an ICCD camera (Andor iStar 320T). 

To locate the conformer-specific electronic origins, the PDL wavelength was scanned while the monochromator grating was tuned synchronously to track the corresponding diagonal fluorescence. For off-diagonal excitation scans, the PDL wavelength was scanned while the monochromator grating was fixed near the corresponding electronic origin. Because corresponding vibrational frequencies are similar in the ground and excited electronic states, vibrationally diagonal emission from the excited vibrational levels occur near this origin. DLIF spectra were acquired by fixing the PDL wavelength at a selected excitation resonance and recording the wavelength-dispersed fluorescence with the monochromator and ICCD camera. The PDL wavelength was scanned in increments of 0.1 nm. Signals were accumulated over 100 and 200 laser shots at each excitation wavelength for the electronic-origin and off-diagonal scans, respectively. The origin-excited and off-diagonal DLIF spectra were accumulated over 20,000 and 1,000 laser shots, respectively. Vibrational branching fractions were extracted following the procedures described in Refs. \cite{lao2022sroph,Lao2025} No correction was applied for the wavelength-dependent instrument response because its variation over the relevant spectral range was estimated to change the diagonal branching fractions by less than 1\%.

\subsection{Theoretical methods}

Harmonic Franck–Condon factors (FCFs) were computed using the following scheme.
The ground-state geometry was first optimized with the PBE0 DFT functional and
the def2-TZVPPD basis set.\cite{Adamo_1999, Hellweg_2015} The D3 dispersion
correction was applied in all DFT calculations.\cite{grimme2010consistent}
Harmonic frequencies were then evaluated at the equilibrium geometry, along with
the third and fourth derivatives of the potential required for the subsequent
anharmonic treatment. Excited-state geometries and harmonic frequencies were
obtained for the target excited states using TD-DFT. The ground- and
excited-state Hessians were then used to perform a harmonic FCF calculation with
the ezFCF software.\cite{Gozem_2021} All DFT calculations were carried out in
the Gaussian16 software package.\cite{frisch2016gaussian}

To account for anharmonic effects such as Fermi resonance, the harmonic states
were corrected using the VPT2 Python code.\cite{Boyer_2022} Intensity borrowing
was included by multiplying the harmonic FCF by the expansion coefficient of the
state participating in the resonance. Our approach followed that established in
Ref. \cite{zhu2024extending}.

\renewcommand*{\thetable}{S\arabic{table}}
\setcounter{table}{0}

\begin{table*}
    \caption{Calculated harmonic and anharmonic VPT frequencies of both conformers in the ground state \X.  All vibrational modes are ordered with increasing frequency. All units are in \cm.}
    \centering
    \setlength{\tabcolsep}{14pt}
    \begin{tabular}{ccccccc}
    \hline
    \hline
\multicolumn{1}{c}{\multirow{2}{*}{Modes}} & \multicolumn{1}{c}{\multirow{2}{*}{Symm.}} & \multicolumn{2}{c}{\anti}   && \multicolumn{2}{c}{\syn}    \\ \cline{3-4} \cline{6-7}
\multicolumn{1}{c}{}                       & \multicolumn{1}{c}{}                       & Harm freq. & Anharm. freq. && Harm freq. & Anharm. freq. \\ \hline
$\nu_1$    & A$'$  & 54   & 54   && 55   & 49   \\
$\nu_2$    & A$'$  & 57   & 64   && 58   & 49   \\
$\nu_3$    & A$''$ & 231  & 228  && 231  & 216  \\
$\nu_4$    & A$''$ & 247  & 236  && 248  & 255  \\
$\nu_5$    & A$'$  & 300  & 304  && 297  & 299  \\
$\nu_6$    & A$''$ & 357  & 335  && 356  & 322  \\
$\nu_7$    & A$'$  & 359  & 361  && 360  & 358  \\
$\nu_8$    & A$''$ & 471  & 457  && 468  & 462  \\
$\nu_9$    & A$'$  & 500  & 499  && 504  & 502  \\
$\nu_{10}$ & A$'$  & 535  & 532  && 534  & 529  \\
$\nu_{11}$ & A$'$  & 635  & 628  && 635  & 627  \\
$\nu_{12}$ & A$''$ & 650  & 637  && 649  & 635  \\
$\nu_{13}$ & A$''$ & 705  & 683  && 705  & 843  \\
$\nu_{14}$ & A$''$ & 771  & 762  && 779  & 738  \\
$\nu_{15}$ & A$'$  & 807  & 797  && 808  & 583  \\
$\nu_{16}$ & A$''$ & 858  & 841  && 857  & 238  \\
$\nu_{17}$ & A$''$ & 880  & 860  && 875  & 813  \\
$\nu_{18}$ & A$''$ & 980  & 966  && 988  & 1179 \\
$\nu_{19}$ & A$'$  & 1001 & 985  && 1008 & 952  \\
$\nu_{20}$ & A$'$  & 1016 & 1004 && 1016 & 1047 \\
$\nu_{21}$ & A$'$  & 1106 & 1088 && 1098 & 1081 \\
$\nu_{22}$ & A$'$  & 1171 & 1155 && 1168 & 1166 \\
$\nu_{23}$ & A$'$  & 1180 & 1168 && 1191 & 1207 \\
$\nu_{24}$ & A$'$  & 1248 & 1236 && 1231 & 1188 \\
$\nu_{25}$ & A$'$  & 1316 & 1290 && 1325 & 1262 \\
$\nu_{26}$ & A$'$  & 1364 & 1334 && 1362 & 1422 \\
$\nu_{27}$ & A$'$  & 1378 & 1355 && 1381 & 1334 \\
$\nu_{28}$ & A$'$  & 1519 & 1483 && 1500 & 1476 \\
$\nu_{29}$ & A$'$  & 1532 & 1506 && 1544 & 1616 \\
$\nu_{30}$ & A$'$  & 1640 & 1601 && 1653 & 1609 \\
$\nu_{31}$ & A$'$  & 1677 & 1638 && 1667 & 1621 \\
$\nu_{32}$ & A$'$  & 3174 & 3036 && 3171 & 3034 \\
$\nu_{33}$ & A$'$  & 3191 & 3057 && 3184 & 3048 \\
$\nu_{34}$ & A$'$  & 3208 & 3069 && 3205 & 3064 \\
$\nu_{35}$ & A$'$  & 3211 & 3079 && 3222 & 3070 \\
$\nu_{36}$ & A$'$  & 3887 & 3702 && 3887 & 3696  \\ 
\hline
\hline
\label{tab:theo-freqs}
\end{tabular}%
\end{table*}

\begin{table*}
\centering
\caption{Calculated harmonic and anharmonic (VPT) frequencies and Franck-Condon factors (FCF $\gtrsim$ 10$^{-4}$) of vibrational decays from both \AtoX~and \BtoX~transitions for the \anti~and \syn~conformers.}
\label{tab:theo-fcf}
\setlength{\tabcolsep}{8pt}
\begin{tabular}{ccccccccccc}
\hline\hline
\multicolumn{5}{c}{\textit{anti} A--X} 
& \multicolumn{5}{c}{\textit{anti} B--X} \\
\cline{1-5} \cline{7-11}
Final& Harm.  & Anharm.  & Harm.  & Anharm. && Final  & Harm. & Anharm.  & Harm.  & Anharm. \\
states & freq. & freq. & FCF & FCF && states & freq. & freq. & FCF & FCF\\
\hline
0                  & 0    & 0    & 0.9587       & 0.9587  & & 0                & 0    & 0    & 0.9740       & 0.9740 \\
$5_1$              & 300  & 303  & 0.0285       & 0.0179  & & $5_1$            & 300  & 303  & 0.0178       & 0.0112 \\
$2_1 4_1$          & 304  & 289  & $<10^{-4}$   & 0.0106  & & $2_1 4_1$        & 304  & 289  & $<10^{-4}$   & 0.0066 \\
$11_1$             & 635  & 628  & 0.0027       & 0.0027  & & $11_1$           & 635  & 628  & 0.0029       & 0.0029 \\
$15_1$             & 807  & 796  & 0.0017       & 0.0017  & & $15_1$           & 807  & 796  & 0.0012       & 0.0012 \\
$7_1$              & 359  & 359  & 0.0027       & 0.0017  & & $7_1$            & 359  & 359  & 0.0012       & 0.0007 \\
$1_1 5_1$          & 355  & 342  & 0.0001       & 0.0009  & & $1_1 5_1$        & 355  & 342  & $<10^{-4}$   & 0.0004 \\
$1_1$              & 54   & 52   & 0.0005       & 0.0005  & & $19_1$           & 1001 & 984  & 0.0003       & 0.0002 \\
$27_1$             & 1378 & 1351 & 0.0008       & 0.0004  & & $1_1$            & 54   & 52   & 0.0002       & 0.0002 \\
$24_1$             & 1248 & 1227 & 0.0004       & 0.0004  & & $24_1$           & 1248 & 1227 & 0.0002       & 0.0002 \\
$22_1$             & 1171 & 1152 & 0.0004       & 0.0004  & & $27_1$           & 1378 & 1351 & 0.0003       & 0.0002 \\
$19_1$             & 1001 & 984  & 0.0005       & 0.0003  & & $4_1 7_1 14_1$   & 1377 & 1355 & $<10^{-4}$   & 0.0001 \\
$4_1 7_1 14_1$     & 1377 & 1355 & $<10^{-4}$   & 0.0003  & & $12_1 17_1$      & 1530 & 1492 & $<10^{-4}$   & 0.0001 \\
$1_1 2_1 4_1$      & 359  & 355  & $<10^{-4}$   & 0.0003  & & $29_1$           & 1532 & 1505 & 0.0003       & 0.0001 \\
$26_1$             & 1364 & 1332 & 0.0004       & 0.0002  & & $9_1$            & 500  & 498  & 0.0001       & 0.0001 \\
$12_1 17_1$        & 1530 & 1492 & $<10^{-4}$   & 0.0002  & & $20_1$           & 1016 & 1002 & 0.0001       & 0.0001 \\
$23_1$             & 1180 & 1160 & 0.0002       & 0.0002  & & $1_1 2_1 4_1$    & 359  & 355  & $<10^{-4}$   & 0.0001 \\
$29_1$             & 1532 & 1505 & 0.0005       & 0.0002  & & $22_1$           & 1171 & 1152 & 0.0001       & 0.0001 \\
$25_1$             & 1316 & 1286 & 0.0002       & 0.0002  & & $25_1$           & 1316 & 1286 & 0.0001       & 0.0001 \\
$7_1 19_1$         & 1361 & 1343 & $<10^{-4}$   & 0.0002  & & $3_1 14_1$       & 1002 & 987  & $<10^{-4}$   & 0.0001 \\
$31_1$             & 1676 & 1636 & 0.0002       & 0.0001  & & $31_1$           & 1676 & 1636 & 0.0001       & 0.0001 \\
$3_1 14_1$         & 1002 & 987  & $<10^{-4}$   & 0.0001  & & $26_1$           & 1364 & 1332 & 0.0001       & 0.0001 \\
$4_1 11_1 12_1$    & 1532 & 1503 & $<10^{-4}$   & 0.0001  & & $4_1 11_1 12_1$  & 1532 & 1503 & $<10^{-4}$   & 0.0001 \\
\hline
\hline
             &             &        &  &                           &        \\     
             &             &        &  &                           &        \\     
\hline
\hline
\multicolumn{5}{c}{\textit{syn} A--X} 
& \multicolumn{5}{c}{\textit{syn} B--X} \\
\cline{1-5} \cline{7-11}
Final& Harm.  & Anharm.  & Harm.  & Anharm. && Final  & Harm. & Anharm.  & Harm.  & Anharm. \\
states & freq. & freq. & FCF & FCF && states & freq. & freq. & FCF & FCF\\
\hline
0                & 0    & 0    & 0.9467       & 0.9467       & & 0            & 0    & 0    & 0.9746       & 0.9746 \\
$5_1$            & 297  & 289  & 0.0292       & 0.0188       & & $5_1$        & 297  & 289  & 0.0158       & 0.0102 \\
$2_1 4_1$        & 306  & 303  & $<10^{-4}$   & 0.0104       & & $2_1 4_1$    & 306  & 303  & $<10^{-4}$   & 0.0056 \\
$1_1$            & 55   & 53   & 0.0074       & 0.0074       & & $11_1$       & 635  & 627  & 0.0030       & 0.0030 \\
$11_1$           & 635  & 627  & 0.0030       & 0.0030       & & $7_1$        & 360  & 359  & 0.0027       & 0.0015 \\
$7_1$            & 360  & 359  & 0.0051       & 0.0029       & & $15_1$       & 808  & 582  & 0.0012       & 0.0012 \\
$15_1$           & 808  & 582  & 0.0017       & 0.0017       & & $1_1 5_1$    & 352  & 342  & $<10^{-4}$   & 0.0006 \\
$1_1 5_1$        & 352  & 342  & 0.0007       & 0.0017       & & $1_1 2_1 4_1$& 361  & 357  & $<10^{-4}$   & 0.0005 \\
$1_1 2_1 4_1$    & 361  & 357  & $<10^{-4}$   & 0.0012       & & $27_1$       & 1381 & 1334 & 0.0004       & 0.0003 \\
$27_1$           & 1381 & 1334 & 0.0010       & 0.0008       & & $19_1$       & 1008 & 991  & 0.0003       & 0.0003 \\
$19_1$           & 1008 & 991  & 0.0005       & 0.0005       & & $24_1$       & 1231 & 1205 & 0.0003       & 0.0002 \\
$24_1$           & 1231 & 1205 & 0.0009       & 0.0005       & & $23_1$       & 1191 & 1202 & $<10^{-4}$   & 0.0001 \\
$23_1$           & 1191 & 1202 & $<10^{-4}$   & 0.0004       & & $26_1$       & 1362 & 1494 & 0.0001       & 0.0001 \\
$26_1$           & 1362 & 1494 & 0.0002       & 0.0003       & & $20_1$       & 1016 & 1048 & 0.0001       & 0.0001 \\
$29_1$           & 1544 & 1528 & 0.0004       & 0.0002       & & $1_1$        & 55   & 53   & 0.0001       & 0.0001 \\
$10_1 19_1$      & 1543 & 1534 & $<10^{-4}$   & 0.0002       & & $10_1$       & 534  & 529  & 0.0001       & 0.0001 \\
$22_1$           & 1168 & 1165 & 0.0002       & 0.0002       & & $29_1$       & 1544 & 1528 & 0.0002       & 0.0001 \\
$31_1$           & 1667 & 1622 & 0.0002       & 0.0002       & & $10_1 19_1$  & 1543 & 1534 & $<10^{-4}$   & 0.0001 \\
$25_1$           & 1325 & 1308 & 0.0001       & 0.0001       & & $31_1$       & 1667 & 1622 & 0.0001       & 0.0001 \\
\hline\hline
\end{tabular}
\end{table*}

\begin{table*}
\caption{Effective VPT Hamiltonians for the $\nu_5/(\nu_2+\nu_4)$ Fermi dyad in the $\widetilde{X}$ state. The Hamiltonians are given in the basis $\{\nu_5,\nu_2+\nu_4\}$. The diagonal elements are the deperturbed zeroth-order energies, and the off-diagonal elements are the Fermi coupling matrix elements. All values are in cm$^{-1}$. The sign of the off-diagonal coupling depends on the phase convention of the basis states.}
\label{tab:fermi_hamiltonians}
\centering
\setlength{\tabcolsep}{14pt}
\renewcommand{\arraystretch}{1.3} 
\begin{tabular}{cccccc}
\hline
\hline
& \multicolumn{2}{c}{\anti} & \multicolumn{2}{c}{\syn} \\
Basis state & $\nu_5$ & $\nu_2+\nu_4$ & $\nu_5$ & $\nu_2+\nu_4$ \\
\hline
$\nu_5$ & 298.013 & 6.871 & 294.212 & $-6.717$ \\ 
$\nu_2+\nu_4$ & 6.871 & 294.417 & $-6.717$ & 298.240 \\
\hline
\hline
\end{tabular}
\end{table*}

\begin{table*}
\centering
\caption{Energies and zeroth-order compositions of the mixed eigenstates obtained by diagonalizing the VPT Hamiltonians in Table~\ref{tab:fermi_hamiltonians}. The dominant contributions are shown in bold.}
\label{tab:fermi_compositions}
\setlength{\tabcolsep}{14pt} 
\renewcommand{\arraystretch}{1.3} 
\begin{tabular}{c c c cc}
\hline
\hline
Conformer & Component & Energy (cm$^{-1}$) & $\nu_5$ (\%) & $\nu_2+\nu_4$ (\%) \\
\hline
\textit{anti} & Lower & 289.113 & 37 & \textbf{63} \\
& Upper & 303.317 & \textbf{63} & 37 \\
\addlinespace
\textit{syn} & Lower & 289.214 & \textbf{64} & 36 \\
& Upper & 303.238 & 36 & \textbf{64} \\
\hline
\hline
\end{tabular}
\end{table*}

\clearpage

\begin{table*}
\centering
\begin{threeparttable}
\caption{Experimental vibrational branching fractions (VBFs) and calculated VPT frequencies and Franck--Condon factors for the observed vibrational decays in DLIF spectra in Fig. \ref{fig:ca-dlif}. The uncertainties of the experimental frequencies are within 5 \cm. VBF values in parentheses are uncertainties propagated from the fitted peak areas and do not include systematic contributions from the wavelegnth-dependent instrument response or unresolved and unobserved decay channels, as discussed in Ref. \cite{zhu2022functionalizing,lao2022sroph,zhu2024extending,Lao2025}.}
\label{tab:exp_vbr}
\setlength{\tabcolsep}{14pt}    
\renewcommand{\arraystretch}{1.3} 
\begin{tabular}{ccccc}
\hline\hline
\multicolumn{5}{c}{\anti \AtoX} \\
\hline
Final states & Exp. freq. (cm$^{-1}$) & Exp. VBF (uncer.) & VPT freq. (cm$^{-1}$) & VPT FCF \\
\hline
0             & 0   & 0.923 (0.015) & 0   & 0.9587 \\
$1_1$         & 55  & 0.015 (0.001) & 52  & 0.0005 \\
$4_1$         & 260 & 0.007 (0.001) & 248 & $<10^{-4}$     \\
$2_1 4_1$     & 290 & 0.028 (0.002) & 289 & 0.0106 \\
$5_1$         & 304 & 0.022 (0.001) & 303 & 0.0179 \\
$7_1$         & 354 & 0.005 (0.001) & 359 & 0.0017 \\
\hline\hline
\multicolumn{5}{c}{\anti \BtoX} \\
\hline
Final states & Exp. freq. (cm$^{-1}$) & Exp. VBF (uncer.) & VPT freq. (cm$^{-1}$) & VPT FCF \\
\hline
0             & 0   & 0.931 (0.003) & 0   & 0.9740 \\
$1_2$         & 92  & 0.022 (0.004) &     & $<10^{-4}$ \\
$2_1 4_1$     & 288 & 0.025 (0.004) & 289 & 0.0066 \\
$5_1$         & 302 & 0.019 (0.004) & 303 & 0.0112 \\
$7_1$         & 354 & 0.004 (0.004) & 359 & 0.0007 \\
\hline\hline
\multicolumn{5}{c}{\syn \AtoX} \\
\hline
Final states & Exp. freq. (cm$^{-1}$) & Exp. VBF (uncer.) & VPT freq. (cm$^{-1}$) & VPT FCF \\
\hline
0             & 0   & 0.915 (0.028)  & 0   & 0.9467 \\
$1_1$         & 51  & 0.013\tnote{*} (0.003) & 53  & 0.0074 \\
$5_1$         & 286 & 0.034 (0.004)  & 289 & 0.0188 \\
$2_1 4_1$     & 302 & 0.022 (0.003)  & 303 & 0.0104 \\
$1_1 5_1$     & 341 & 0.005 (0.004)  & 342 & 0.0017 \\
$7_1$         & 354 & 0.010 (0.004)  & 359 & 0.0029 \\
\hline\hline
\multicolumn{5}{c}{\syn \BtoX} \\
\hline
Final states & Exp. freq. (cm$^{-1}$) & Exp. VBF (uncer.) & VPT freq. (cm$^{-1}$) & VPT FCF \\
\hline
0             & 0   & 0.912 (0.009) & 0   & 0.9746 \\
$1_2$         & 95  & 0.022 (0.004) &     & $<10^{-4}$ \\
$5_1$         & 287 & 0.031 (0.007) & 289 & 0.0102 \\
$2_1 4_1$     & 303 & 0.020 (0.008) & 303 & 0.0056 \\
$7_1$         & 355 & 0.014 (0.005) & 359 & 0.0015 \\
\hline\hline
\end{tabular}
\begin{tablenotes}
\footnotesize
\centering
\item[*]CaOH contamination leads to an overestimate of the experimental $1_1$ branching fraction. The value reported here was therefore rescaled using the VPT calculated FCF ratio $5_1/1_1$ in \syn~A.
\end{tablenotes}
\end{threeparttable}
\end{table*}

\begin{table*}
\centering
\caption{Calculated harmonic vibrational frequencies and symmetries of the
\textit{anti} and \textit{syn} conformers in the
$\widetilde{A}$ and $\widetilde{B}$ states. All vibrational modes are numbered in order of increasing harmonic frequency. Frequencies are in cm$^{-1}$.}
\label{tab:AB_freq}
\setlength{\tabcolsep}{8pt}
\begin{tabular}{c cc c cc c cc c cc}
\hline
\hline
& \multicolumn{2}{c}{\anti-A}
&& \multicolumn{2}{c}{\anti-B}
&& \multicolumn{2}{c}{\syn-A}
&& \multicolumn{2}{c}{\syn-B} \\
\cline{2-3}
\cline{5-6}
\cline{8-9}
\cline{11-12}
Mode
& Sym. & \shortstack{Freq.}
&& Sym. & \shortstack{Freq.}
&& Sym. & \shortstack{Freq.}
&& Sym. & \shortstack{Freq.} \\
\hline
$\nu_{1}$  & $A'$  & 53   && $A''$ & 56   && $A'$  & 53   && $A''$ & 56   \\
$\nu_{2}$  & $A''$ & 60   && $A'$  & 56   && $A''$ & 60   && $A'$  & 57   \\
$\nu_{3}$  & $A''$ & 231  && $A''$ & 231  && $A''$ & 232  && $A''$ & 231  \\
$\nu_{4}$  & $A''$ & 249  && $A''$ & 246  && $A''$ & 250  && $A''$ & 247  \\
$\nu_{5}$  & $A'$  & 298  && $A'$  & 298  && $A'$  & 296  && $A'$  & 295  \\
$\nu_{6}$  & $A'$  & 359  && $A''$ & 358  && $A'$  & 358  && $A''$ & 355  \\
$\nu_{7}$  & $A''$ & 359  && $A'$  & 360  && $A''$ & 360  && $A'$  & 360  \\
$\nu_{8}$  & $A''$ & 471  && $A''$ & 470  && $A''$ & 469  && $A''$ & 468  \\
$\nu_{9}$  & $A'$  & 500  && $A'$  & 501  && $A'$  & 503  && $A'$  & 504  \\
$\nu_{10}$ & $A'$  & 535  && $A'$  & 535  && $A'$  & 535  && $A'$  & 535  \\
$\nu_{11}$ & $A'$  & 637  && $A'$  & 635  && $A'$  & 637  && $A'$  & 635  \\
$\nu_{12}$ & $A''$ & 649  && $A''$ & 648  && $A''$ & 648  && $A''$ & 647  \\
$\nu_{13}$ & $A''$ & 705  && $A''$ & 703  && $A''$ & 705  && $A''$ & 703  \\
$\nu_{14}$ & $A''$ & 772  && $A''$ & 772  && $A''$ & 780  && $A''$ & 780  \\
$\nu_{15}$ & $A'$  & 807  && $A'$  & 807  && $A'$  & 808  && $A'$  & 807  \\
$\nu_{16}$ & $A''$ & 860  && $A''$ & 859  && $A''$ & 859  && $A''$ & 856  \\
$\nu_{17}$ & $A''$ & 882  && $A''$ & 878  && $A''$ & 878  && $A''$ & 874  \\
$\nu_{18}$ & $A''$ & 981  && $A''$ & 981  && $A''$ & 991  && $A''$ & 989  \\
$\nu_{19}$ & $A'$  & 1002 && $A'$  & 1001 && $A'$  & 1009 && $A'$  & 1008 \\
$\nu_{20}$ & $A'$  & 1016 && $A'$  & 1016 && $A'$  & 1016 && $A'$  & 1016 \\
$\nu_{21}$ & $A'$  & 1106 && $A'$  & 1107 && $A'$  & 1098 && $A'$  & 1099 \\
$\nu_{22}$ & $A'$  & 1170 && $A'$  & 1170 && $A'$  & 1168 && $A'$  & 1168 \\
$\nu_{23}$ & $A'$  & 1180 && $A'$  & 1180 && $A'$  & 1191 && $A'$  & 1191 \\
$\nu_{24}$ & $A'$  & 1246 && $A'$  & 1247 && $A'$  & 1229 && $A'$  & 1229 \\
$\nu_{25}$ & $A'$  & 1316 && $A'$  & 1316 && $A'$  & 1325 && $A'$  & 1325 \\
$\nu_{26}$ & $A'$  & 1362 && $A'$  & 1363 && $A'$  & 1363 && $A'$  & 1362 \\
$\nu_{27}$ & $A'$  & 1374 && $A'$  & 1375 && $A'$  & 1376 && $A'$  & 1377 \\
$\nu_{28}$ & $A'$  & 1519 && $A'$  & 1519 && $A'$  & 1499 && $A'$  & 1500 \\
$\nu_{29}$ & $A'$  & 1530 && $A'$  & 1530 && $A'$  & 1542 && $A'$  & 1542 \\
$\nu_{30}$ & $A'$  & 1642 && $A'$  & 1642 && $A'$  & 1655 && $A'$  & 1654 \\
$\nu_{31}$ & $A'$  & 1676 && $A'$  & 1676 && $A'$  & 1666 && $A'$  & 1666 \\
$\nu_{32}$ & $A'$  & 3175 && $A'$  & 3176 && $A'$  & 3171 && $A'$  & 3171 \\
$\nu_{33}$ & $A'$  & 3192 && $A'$  & 3192 && $A'$  & 3185 && $A'$  & 3185 \\
$\nu_{34}$ & $A'$  & 3208 && $A'$  & 3208 && $A'$  & 3205 && $A'$  & 3205 \\
$\nu_{35}$ & $A'$  & 3210 && $A'$  & 3210 && $A'$  & 3222 && $A'$  & 3223 \\
$\nu_{36}$ & $A'$  & 3886 && $A'$  & 3886 && $A'$  & 3886 && $A'$  & 3887 \\
\hline
\hline
\end{tabular}
\end{table*}

\begin{table*}
\centering
\caption{Assignments of the off-diagonal vibrational resonances observed in the \A~ and \B~states of both conformers in Fig. \ref{fig:ca-pes-ex}d. Excitation energies are reported relative to the corresponding vibrational origins. Tentative assignments are made by comparing the measured energy offsets, $\Delta E$, with calculated harmonic frequencies (Table \ref{tab:AB_freq}) and FCFs (Table \ref{tab:theo-fcf}). Bold values indicate the energy offsets associated with the dominant conformer-state contributions and used for the assignments. The final column identifies the DLIF spectra recorded at selected excitation wavelengths.}
\label{tab:ex_peaks_assign}
\setlength{\tabcolsep}{7pt}
\begin{tabular}{cccccccc}
\hline
\hline
Observed peaks & Excitation  & \multicolumn{4}{c}{$\Delta E$ (\cm) relative to} & Assignments & DLIF  \\
\cline{3-6}
in Fig. \ref{fig:ca-pes-ex}d & wavelength (nm) & \anti-A0 & \anti-B0 & \syn-A0 & \syn-B0 &  & spectra  \\
\hline
${}^{A}1/{}^{B}1$        & 605.65 & 266           & 146           & \textbf{292}  & 120           & \syn A, $2^1 4^1$                                  &         \\
${}^{A}2$                & 605.05 & 282           & 163           & \textbf{309}  & 136           & \syn A, $5^1$                                      & Fig.~4  \\
${}^{A}3/{}^{B}2$        & 604.75 & \textbf{291}  & 171           & 317           & 144           & \anti A, $2^1 4^1$                                 & Fig.~S3 \\
${}^{A}4$                & 604.15 & \textbf{307}  & 187           & 333           & 161           & \anti A, $5^1$                                     & Fig.~4  \\
${}^{A}5/{}^{B}3$        & 603.35 & 329           & 209           & \textbf{355}  & 183           & \syn A, $7^1$                                      & Fig.~S3 \\
${}^{A}6/{}^{B}4$        & 602.75 & \textbf{346}  & 226           & 372           & 199           & \anti A, $1^1 5^1$                                 & Fig.~S3 \\
${}^{A}7/{}^{B}5$        & 602.45 & \textbf{354}  & 234           & 380           & 207           & \anti A, $7^1$                                     & Fig.~S3 \\
${}^{A}8/{}^{B}6$        & 601.15 & 390           & 270           & 416           & \textbf{243}  & \syn B, $4^1$                                      & Fig.~S3 \\
${}^{A}9/{}^{B}7$        & 600.35 & \textbf{412}  & \textbf{292}  & 438           & 265           & \anti A, $1^1 7^1$; \anti B, $2^1 4^1$            & Fig.~S3 \\
${}^{A}10/{}^{B}8$       & 599.85 & 426           & \textbf{306}  & 452           & 279           & \anti B, $5^1$                                     & Fig.~S3 \\
${}^{A}11/{}^{B}9$       & 599.45 & 437           & 317           & 463           & \textbf{290}  & \syn B, $2^1 4^1$                                  &         \\
${}^{A}12/{}^{B}10$      & 598.85 & 454           & 334           & 480           & \textbf{307}  & \syn B, $5^1$                                      & Fig.~S3 \\
${}^{A}13/{}^{B}11$      & 598.25 & 470           & \textbf{351}  & 497           & 324           & \anti B, $1^1 5^1$                                 &         \\
${}^{A}14/{}^{B}12$      & 598.05 & 476           & \textbf{356}  & 502           & 329           & \anti B, $7^1$                                     & Fig.~S3 \\
${}^{A}15$               & 597.55 & \textbf{490}  & 370           & 516           & 343           & \anti A, $9^1$                                     &         \\
${}^{B}13$               & 597.45 & 493           & 373           & 519           & \textbf{346}  & \syn B, $1^1 5^1$                                  &         \\
${}^{A}16/{}^{B}14$      & 597.05 & 504           & 384           & 530           & \textbf{357}  & \syn B, $7^1$                                      & Fig.~S3 \\
${}^{B}15$               & 595.95 & 535           & \textbf{415}  & 561           & 388           & \anti B, $1^1 7^1$                                 &         \\
${}^{A}17/{}^{B}16$      & 595.15 & 557           & 438           & \textbf{584}  & \textbf{411}  & \syn A, $2^1 4^1 5^1$; \syn B, $1^1 7^1$          & Fig.~S3 \\
${}^{A}18$               & 594.65 & 572           & 452           & \textbf{598}  & 425           & \syn A, $5^2$                                      &         \\
${}^{A}19$               & 594.25 & \textbf{583}  & 463           & 609           & 436           & \anti A, $2^1 4^1 5^1$                             &         \\
${}^{A}20$               & 593.85 & \textbf{594}  & 475           & 621           & 448           & \anti A, $5^2$                                     &         \\
${}^{A}21/{}^{B}17$      & 593.55 & 603           & \textbf{483}  & \textbf{629}  & 456           & \syn A, $11^1$; \anti B, $9^1$                     & Fig.~4  \\
${}^{A}22/{}^{B}18$      & 592.65 & \textbf{628}  & 509           & 655           & 482           & \anti A, $11^1$                                    & Fig.~4  \\
${}^{B}19$               & 592.05 & 645           & \textbf{526}  & 672           & 499           & \anti B, $3^1 5^1$                                 &         \\
${}^{A}23/{}^{B}20$      & 590.05 & 703           & \textbf{583}  & 729           & 556           & \anti B, $2^1 4^1 5^1$                             &         \\
${}^{A}24/{}^{B}21$      & 589.55 & 717           & \textbf{597}  & 743           & 571           & \anti B, $5^2$                                     &         \\
${}^{A}25$               & 588.85 & 737           & 618           & \textbf{763}  & 591           & \syn A, $1^1 7^2$                                  &         \\
${}^{A}26/{}^{B}22$      & 588.45 & 749           & \textbf{629}  & 775           & 602           & \anti B, $11^1$                                    &         \\
${}^{A}27$               & 587.95 & 763           & 644           & \textbf{789}  & 617           & \syn A, $15^1$                                     &         \\
${}^{A}28/{}^{B}23$      & 587.55 & 775           & 655           & \textbf{801}  & 628           & \syn A, $5^1 9^1$                                  &         \\
${}^{A}29$               & 587.05 & \textbf{789}  & 670           & 816           & 643           & \anti A, $15^11$                                    &         \\
${}^{B}24$               & 586.65 & 801           & \textbf{681}  & 827           & 654           & \anti B, $1^1 11^1$                                &         \\
${}^{A}30$               & 585.25 & \textbf{842}  & 722           & 868           & 695           & \anti A, $1^1 5^1 9^1$                             &         \\
${}^{A}31$               & 584.85 & \textbf{853}  & 734           & 880           & 707           & \anti A, $7^1 9^1$                                 &         \\
${}^{A}32/{}^{B}25$      & 584.25 & \textbf{871}  & 751           & 897           & 724           & \anti A, $7^1 10^1$                                &         \\
${}^{A}33/{}^{B}26$      & 582.95 & \textbf{909}  & \textbf{789}  & 935           & 763           & \anti B, $15^1$; \anti A, $5^3$                    & Fig.~4  \\
${}^{B}27$               & 582.05 & 936           & 816           & 962           & \textbf{789}  & \syn B, $15^1$                                     & Fig.~4  \\
${}^{A}34/{}^{B}28$      & 580.85 & \textbf{971}  & \textbf{851}  & 997           & 825           & \anti A, $7^1 11^1$; \anti B, $7^1 9^1$           & Fig.~S3 \\
${}^{A}35$               & 577.35 & \textbf{1075} & 956           & 1102          & 929           & \anti A, $5^1 15^1$                                & Fig.~S3 \\
${}^{B}30$               & 575.95 & 1118          & \textbf{998}  & \textbf{1144} & 971           & \anti B, $20^1$; \syn A, $22^1$                  & Fig.~S3 \\
${}^{A}36$               & 575.15 & \textbf{1142} & 1022          & \textbf{1168} & 995           & \anti A, $22^1$; \syn A, $23^1$                    & Fig.~S3 \\
\hline
\hline
\end{tabular}
\end{table*}

\renewcommand*{\thefigure}{S\arabic{figure}}
\setcounter{figure}{0} 
\begin{figure*}
    \centering
    \includegraphics[scale=1]{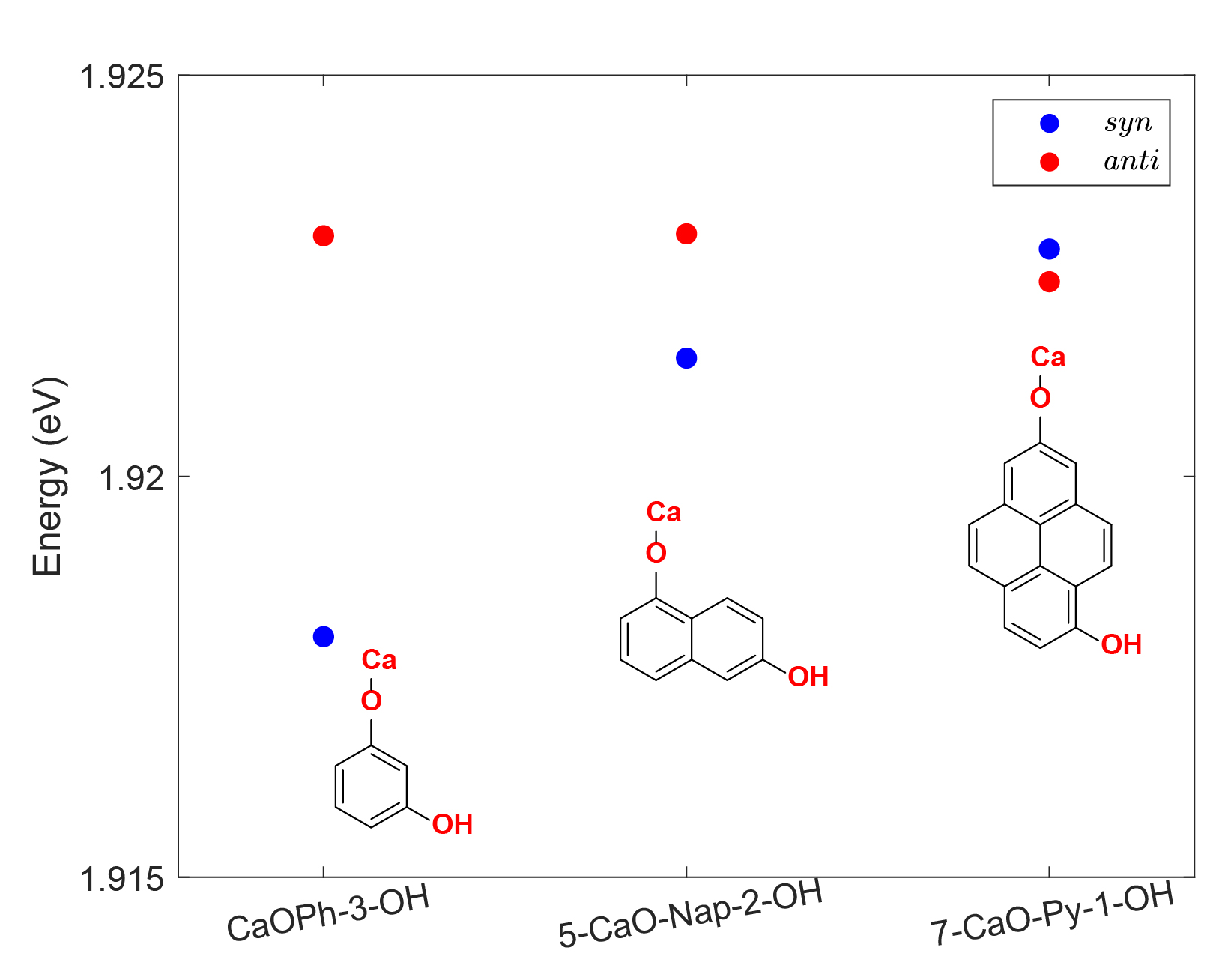}
    \caption{Computed TDDFT vertical excitation energies for a series of
    increasingly large conjugated molecules. The blue and red points correspond
    to the excitation energy of the $\tilde{A}$ state for the syn- and anti-
    conformers, accordingly.} 
    \label{fig:TDDFT-fig}
\end{figure*}

\begin{figure*}
    \centering
    \includegraphics[scale=0.35]{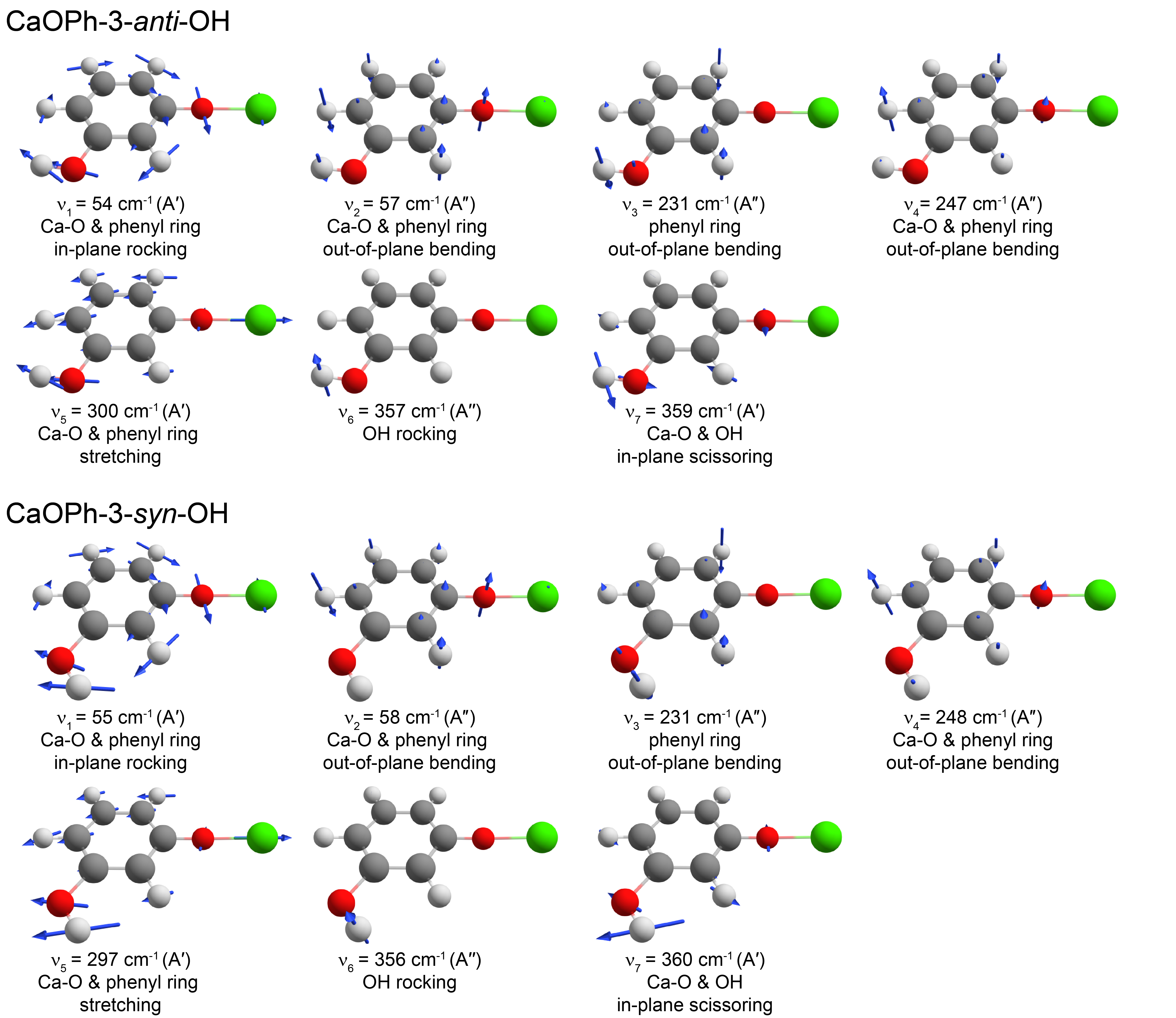}
    \caption{Vibrational displacements of the lowest seven fundamental vibrational modes for both conformers. The calculated harmonic frequencies and symmetries are given. }
    \label{fig:vib-modes}
\end{figure*}

\begin{figure*}
    \centering
    \includegraphics[scale=0.5]{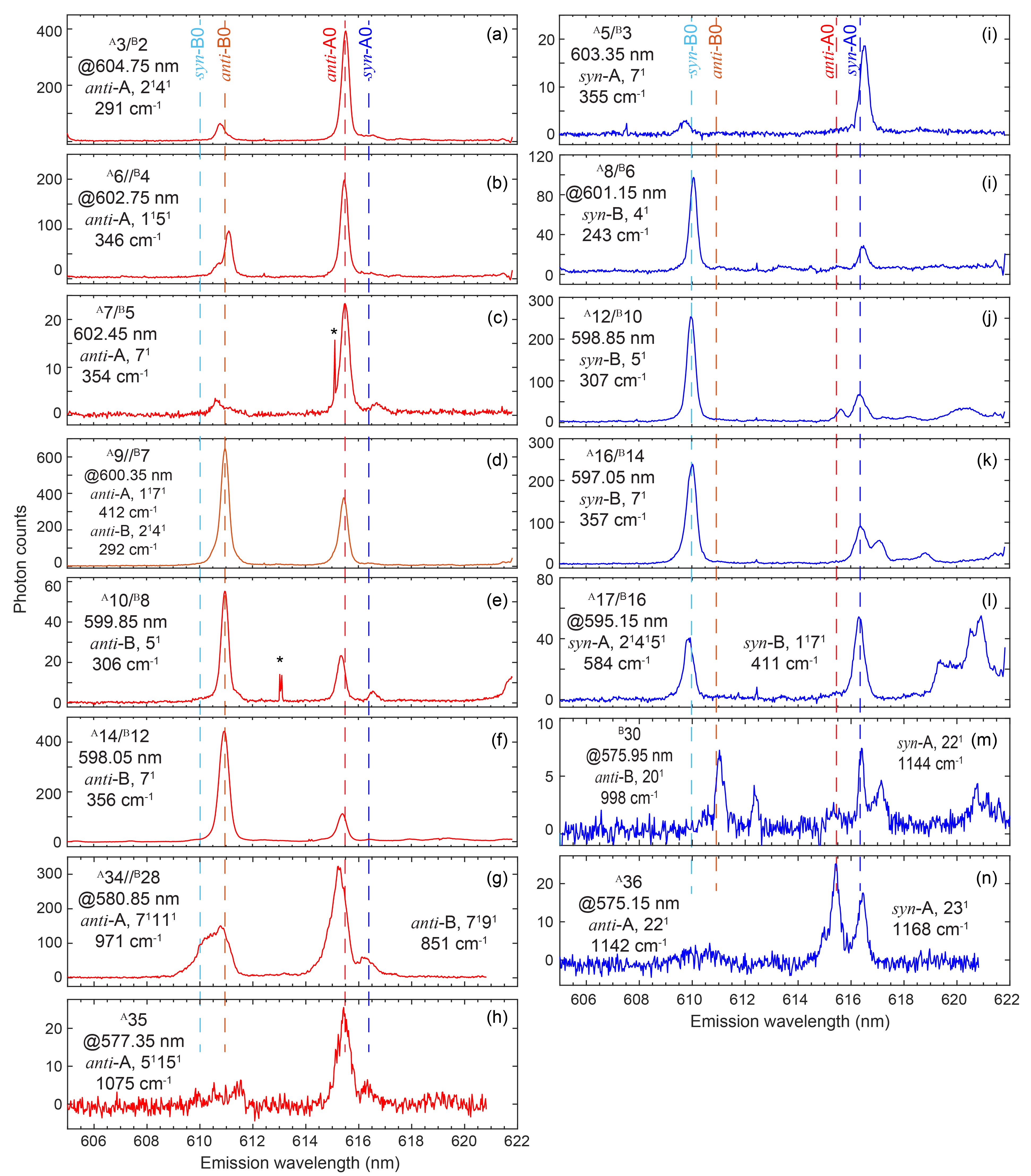}
    \caption{The remaining fifteen DLIF spectra showing the excitations of selected vibrational resonances in Fig. \ref{fig:ca-pes-ex}d. Each spectrum is dominated by vibrationally diagonal emission associated with one of the four conformer-specific electronic origins: \anti-A0, \syn-A0, \anti-B0, or \syn-B0. The excitation energy offsets relative to the corresponding electronic origins and the corresponding assignments are indicated in each figure. Assignments of the excitation resonances are obtained by comparing those offsets with the calculated vibrational frequencies (Table. \ref{tab:AB_freq}). Asterisks denote experimental artifacts.}
    \label{fig:SI-Ca-emission}
\end{figure*}

\end{document}